\documentclass[letter, amsfonts, amssymb, amsmath, reprint, showkeys, twoside, longbibliography]{revtex4-1}

\usepackage[english]{babel}
\usepackage[utf8]{inputenc}
\usepackage[colorinlistoftodos, color=green!40, prependcaption]{todonotes}
\usepackage{amsthm}
\usepackage{mathtools}
\usepackage{physics}
\usepackage{xcolor}
\usepackage{graphicx}
\usepackage[left=23mm,right=13mm,top=35mm,columnsep=15pt]{geometry} 
\usepackage{adjustbox}
\usepackage{placeins}
\usepackage[T1]{fontenc}
\usepackage{lipsum}
\usepackage{csquotes}

\usepackage[%
  colorlinks=true,
  urlcolor=blue,
  linkcolor=blue,
  citecolor=blue
]{hyperref}
\usepackage{etoolbox}
\usepackage{breqn}

\newcommand{\be}{\begin{equation}}
\newcommand{\ee}{\end{equation}}
\newcommand{\bea}{\begin{eqnarray}}
\newcommand{\eea}{\end{eqnarray}}
\newcommand{\bei}{\begin{itemize}}
\newcommand{\eei}{\end{itemize}}

\begin{document}
\title{Intersection Numbers in Quantum Mechanics and Field Theory}

\author{Sergio L. Cacciatori}
\email{sergio.cacciatori@uninsubria.it}
    \affiliation{\uninsubria \\ \miinfn}

\author{Pierpaolo Mastrolia}
\email{pierpaolo.mastrolia@unipd.it}
    \affiliation{\unipd \\ \pdinfn}

\newcommand{\unipd}{Dipartimento di Fisica e Astronomia, Universit\`a degli Studi di Padova,
Via Marzolo 8, I-35131 Padova, Italy}

\newcommand{\pdinfn}{INFN, sezione di Padova,
Via Marzolo 8, I-35131 Padova, Italy}

\newcommand{\uninsubria}{Department of Science and High Technology, Universit\`a  dell'Insubria, Via Valleggio 11, 22100, Como, Italy}

\newcommand{\miinfn}{INFN, sezione di Milano, Via Celoria 16, 20133, Milano, Italy}

\date{\today} 

\begin{abstract} 
By elaborating on the recent progress made in the area of Feynman integrals, we apply the intersection theory for twisted de Rham cohomologies to simple integrals involving orthogonal polynomials, matrix elements of operators in Quantum Mechanics and Green's functions in Field Theory, showing that the algebraic identities they obey are related to the decomposition of twisted cocycles within cohomology groups, 
and which, therefore, can be derived by means of intersection numbers.
Our investigation suggests an algebraic approach generically applicable to the study 
of higher-order moments of probability distributions, where 
the dimension of the cohomology groups corresponds to the number of independent moments;
the intersection numbers for twisted cocycles can be used to derive linear and quadratic relations among them.
Our study offers additional evidence of the intertwinement between physics, geometry, and statistics.
\end{abstract}


\maketitle

\section{Introduction} \label{sec:intro}

Computing integrals is routine in any scientific ambit: expectation values in Quantum Mechanics, Feynman integrals in Quantum Field Theory, higher momenta in Statistics are just a few paradigmatic examples out of a plethora of cases. 
Stokes' theorem represents a first step toward a unifying vision of integrals evaluation as a whole: 
when it is possible to look at them as representing fluxes of closed differential forms through (hyper)surfaces, it tells us that such integrals are invariant under deformations either of the integrand by exact forms, or the contour by boundary terms. 
This gives rise to the de Rham theory of cohomology, and its generalizations, as its twisted version that allows to include singular differential forms.
For these reasons, within cohomology theories, the analytic properties of functions are tight to the algebraic properties of the elements appearing in the corresponding integral representations (forms and contours), which, in turn, are determined by the geometry (holes and singularities) of their existence domains.
\par
The linearity of integral calculus makes not surprising that (regulated bounded) integrals form a vector space structure.
The intersection theory of twisted de Rham cohomology~\cite{aomoto2011theory,yoshida2013hypergeometric,MANA:MANA19941660122,MANA:MANA19941680111,matsumoto1994,cho1995,AOMOTO1997119,matsumoto1998,adolphson_sperber_1997,Mimachi2003,Ohara98intersectionnumbers,Mimachi2004,OST2003,aomoto2011theory,Goto:2013Laur,Yoshiaki-GOTO2015203,gotomatsumoto2015,matsumoto2018relative,matsubaraheo2019algorithm,Goto2022homology,Matsubara-Heo:2020lzo,Matsubara-Heo:2021dtm} offers the proper mathematical framework to characterize it and to establish linear and quadratic relations involving the integrals. These relations emerge from the {\it intersection numbers} of either regulated/loaded contours, known as {\it twisted cycles}, or differential forms, known as {\it twisted cocycles}. 
They belong to two isomorphic vector spaces, known as {\it homology groups} and {\it cohomology groups}, which are generated by {\it bases} of twisted cycles and cocycles, respectively.
Homology and cohomology groups are isomorphic, so they have the same dimension, whose value depends on the geometry of the variety associated with a regulating multivalued function that appears in the integral representation, known as the {\it twist}.
The geometric properties of this function, namely its zeroes and the critical points, determine the algebraic and analytic properties of the integrals.
\par
More precisely, the cohomology groups associated with the class of integrals we deal with in this study 
are finite dimensional vector spaces in which the intersection products act as inner products, namely bilinear pairings between the group elements. 
Within the homology group the inner product is realized by the so-called {\it topological intersection number} between cycles, and can be viewed as the generalization of 
the number of intersections of two curves in a plane; 
within the cohomology group, it is called {\it intersection number} of twisted cocycles, and is a key operation in this communication - its visualization is less intuitive.
Therefore linear and quadratic relations, as well as the differential and finite difference equation they obey, emerge from the decomposition of differential forms 
or analogously, from the decomposition of contours, in terms of a set of generators.
\\
De Rham's theory has been applied to study the vector space structures of several types of functions 
widely appearing in mathematics and physics, such as
Aomoto-Gel'fand and Gauss hypergeometric integrals,  
Gel'fand-Kapranov-Zelevinsky systems.
Euler-Mellin integrals.
as well as Feynman integrals.
In the latter case, it allowed to reformulate the problem of the decomposition of Feynman integrals in terms of master integrals, usually achieved through the integration-by-parts (IBP) algorithm \cite{Chetyrkin:1981qh,Laporta:2001dd}, 
and to address it as a well-posed linear algebra problem, dealing with the decomposition of generic vectors in a basis of a vector space \cite{Mastrolia:2018uzb,Frellesvig:2019kgj,Mizera:2019gea,Mizera:2019vvs,Frellesvig:2019uqt,Frellesvig:2020qot,Chestnov:2022alh,Chestnov:2022xsy}. In this fashion, the system-solving algorithm characterizing the IBP algorithm is replaced by projections via intersection numbers. 
\par
In this work, we elaborate on the role of the intersection numbers of twisted cocycles, 
which is an operation based on the interplay of Stokes' and Residues' theorems, originally introduced in a purely mathematical context, for the study of hypergeometric integrals \cite{cho1995}, and whose importance in physics has been made manifest more recently in the context of scattering amplitudes and Feynman integrals evaluation \cite{Mizera:2016jhj,Mizera:2017rqa,Mastrolia:2018uzb,Frellesvig:2019kgj,Mizera:2019gea,Mizera:2019vvs,Frellesvig:2019uqt,Frellesvig:2020qot}, 
recently discussed also  in~\cite{Mizera:2019ose,Cacciatori:2021nli,Frellesvig:2021vem,Weinzierl:2022eaz,Matsumoto:2022Va,Mastrolia:2022tww,Mandal:2022vok} 
(see also \footnote{ Proceedings of conference, {\it MathemAmplitudes 2019: Intersection Theory \& Feynman Integrals}, Padova, available at {\tt https://pos.sissa.it/383/} .}).
These applications triggered the development of refined methods for the evaluation of intersection numbers 
\cite{Mizera:2019gea,Frellesvig:2019uqt,Weinzierl:2020xyy,Caron-Huot:2021xqj,Caron-Huot:2021iev,Chestnov:2022alh,Chestnov:2022xsy}, 
and other interesting physics applications \cite{Kaderli:2019dny,Kalyanapuram:2020vil,Ma:2021cxg,Weinzierl:2020nhw,Gasparotto:2022mmp,Chen:2022lzr,Giroux:2022wav}. 

Our current study aims at providing additional new proof strongly enforcing the evidence
of the role of the intersection theory of de Rham twisted cohomologies in the context of fundamental physics and its connection to statistics, by showing that the integral decomposition via intersection numbers 
can be applied to the algebra of orthogonal polynomials,  
to the computation of matrix elements in Quantum Mechanics, 
and to the algebraic properties of Green's function and Wick's theorem in Quantum Field Theory, 
which are, more generally, related to the higher-order moments of distributions and Isserlis' theorem in Probability Theory.

\section{Integrals and Twisted Cohomology groups} \label{sec:integrals}

Within the intersection theory of twisted de Rham cohomology, 
any regulated bounded integral, called {\it twisted period integral},
\bea
\int_{\Gamma} u \, \varphi
\label{eq:def:twistedintegral}
\eea
is defined as a {\it pairing} of the twisted cocycle
$ 
\varphi \equiv {\hat \varphi} \, d^n z \ ,
$ 
that is an element of the {\it de Rham} $n$-th {\it cohomology group} $H^n_\omega$,
and the twisted cycle $(\Gamma, u)$, 
that is an element of the {\it de Rham} $n$-th {\it homology group} $H_n^\omega$, 
characterized by the property that $u$ vanishes on the boundary of $\Gamma$, i.e. $u(\partial \Gamma)=0$.
In the current study, ${\hat \varphi}$ and the twist $u$ are respectively meromorphic and multivalued functions of the complex integration variables $z_1, \ldots, z_n$ \footnote{
Generally, $u \equiv Q \prod_j P_j^{\rho_j}$, where $P_j$ and $Q$ are meromorphic with allowed singularities at infinity, and $\rho_j$ are a generic exponents.},
whereas the connection $\omega$, is defined as,
\bea
\omega \equiv d \ln(u) = 
\sum_{i=1}^n {\hat \omega}_i \, d z_i \ , \quad {\rm with \ }
{\hat \omega}_i \equiv \partial_i \ln(u) \ .
\eea
Let us observe that $H^n_\omega$ is the quotient space of closed $n$-forms
$ \{\varphi  : \nabla_\omega \varphi = 0 \}$
modulo exact forms 
$ \{ \varphi \, : \ \ \varphi = \nabla_\omega \varphi_{n-1} \}$, 
where $\varphi_{n-1}$ is a differential $(n-1)$-form,
with respect to the {\it covariant derivative} $\nabla_\omega$, defined as,
\bea
\nabla_\omega \equiv  d + \omega \wedge = u^{-1} \cdot d \cdot u \ . 
\eea
Therefore, $H^n_\omega$ is the space of the differential $n$-forms which differ by terms that vanish upon integration, 
$\varphi \sim \varphi + \nabla_\omega \, \varphi_{n-1}$, 
hence leaving the integral invariant, 
\bea
\int_{\Gamma} u \, \varphi = 
\int_{\Gamma} u \, (\varphi + \nabla_\omega \, \varphi_{n-1}) \ ,
\eea
because 
\bea 
  \int_{\Gamma} d (u \, \varphi_{n-1}) 
= \int_{\Gamma} u \, \nabla_\omega \, \varphi_{n-1} = 0 \ .
\eea
The vector spaces $H_n^\omega$ and $H^n_\omega$ are isomorphic, hence they have the same dimensions,
\bea
\nu \equiv {\rm dim}(H^n_{\omega}) \ ,
\eea

\noindent
which generically corresponds to the number of critical points of Morse's {\it height} function $ \ln(u) $ 
\footnote{
The known isomorphism of the de Rham cohomology groups offers several ways of computing its dimension:
within algebraic geometry,
by counting the number of {\it critical points} of $\ln u $, ${\hat \omega}_i=0$ $i=1,\ldots,n$ \cite{Lee:2013hzt}; 
or, the number of solutions of the zero-dimensional system with the Shape Lemma \cite{Frellesvig:2020qot}, 
as the dimension of the quotient space of complex polynomials in the variables $z_1,\ldots,z_n$, 
modulo the ideal $I=\langle {\hat \omega}_1, \ldots, {\hat \omega}_n \rangle$, i.e. $\nu = {\rm dim}({\bf C}[\{z_1,\ldots,z_n\}]/I)$;
within  differential topology,
from the Euler characteristic $\chi( {\mathbb P}_{\omega})$ 
of the projective variety ${\mathbb P}_{\omega}$ 
generated by the poles of $\omega$, i.e. $\nu = (-1)^n (n + 1 - \chi( {\mathbb P}_{\omega} ))$ \cite{Frellesvig:2019uqt,Bitoun:2017nre};
within algebraic combinatorics, 
from the mixed volume of Newton polyhedra,
by using Bernstein–Kushnirenko theorem
\cite{Adolphson-1994};
within differential geometry,
by computing the rank of Macaulay matrices of ${\cal D}$-modules \cite{Chestnov:2022alh};
within statistics, 
by computing the maximum likelihood degree (of toric variety) \cite{10.2307/40067993,Agostini:2021rze}
-- to list a few.
}.

\subsection{Decomposition and Intersection Numbers}

Several problems in mathematics and physics \cite{Schwarz:2008sa} 
may require the evaluation of integrals of the type (\ref{eq:def:twistedintegral}),
which can be generically tackled within a two-step algorithm: 
\bei
\item[1.]
the decomposition in terms of an independent set of integrals, 
say $\{ E_i \}_{i=1}^\nu \, $,
\bea
\int_{\Gamma} u \, \varphi = 
\sum_{i=1}^\nu c_i \, E_i \ ,
%
\label{eq:mintdecomposition}
\eea
where $ E_i $ are also twisted period integrals,
\bea
E_i \equiv \int_{\Gamma} u \, e_{i} \ , \qquad i=1, \ldots, \nu \, .
\eea
hereby called {\it master integrals}, by borrowing the terminology from Feynman integral calculus \cite{Chetyrkin:1981qh,Laporta:2001dd};

\item[2.] the evaluation of master integrals: by means of direct integration, when possible, or by solving the 
linear systems of differential and difference equations they obey \cite{Barucchi:1973zm,KOTIKOV1991158,Remiddi:1997ny,Gehrmann:1999as,Laporta:2001dd,Laporta:2003jz,Tarasov:1996br}, 
generically written as,
\bea
{\cal F}\Big[ E_i \Big] = {\bf F}_{ij} \, E_j \ ,
\eea
where ${\cal F}$ stands for a (differential-) operator acting on the master integrals,
and ${\bf F}$ stands for a linear operator (a matrix) that is built by exploiting the linear relations of the type (\ref{eq:mintdecomposition}). 
\eei

Intersection theory for twisted de Rham cohomologies offers a new perspective and 
a novel solution to the integral decomposition problem, 
which can be addressed as a linear decomposition problem in vector space, and, therefore, solved by projecting of the twisted
cocycle $\varphi$ on a basis of twisted cocycles within the cohomology group $H_\omega^n$. 

The cocycle projection follows from the {\it master decomposition formula} \cite{Mastrolia:2018uzb,Frellesvig:2019kgj,Mizera:2019gea,Mizera:2019vvs,Frellesvig:2019uqt,Frellesvig:2020qot},
\bea 
\label{eq:masterdecoformula}
\varphi = \sum_{i=1}^\nu c_i \, e_{i} \ , \quad {\rm with}\quad 
c_i = \sum_{j=1}^\nu \langle \varphi \, , h_j \rangle \, {\bf C}^{-1}_{ji}  \ ,
\eea
where:
$\{ e_i \}_{i=1}^\nu$ is a basis of twisted cocycles, generating $H^n_{\omega}$;
$\{ h_i \}_{i=1}^\nu$ is a basis of the dual space
$H^n_{-\omega}$ \footnote{
The values of the coefficients $c_i$ in (\ref{eq:masterdecoformula}) do not depend on the actual choice of the dual basis 
$\{ h_i \}_{i=1}^\nu$ \cite{Frellesvig:2019kgj} 
}
; 
and ${\bf C}$ is their {\it intersection matrix},
\bea
 {\bf C}_{ij} \equiv \langle e_i  \, , h_j \rangle \ .
\eea

The master decomposition formula (\ref{eq:masterdecoformula}) relates in the integral decomposition to the decomposition of twisted cocycles, and shows that the coefficients $c_i$ that appear in (\ref{eq:mintdecomposition}) can be computed by means of the {\it cohomology intersection numbers} 
$\langle \varphi_L \, , \varphi_R \rangle$ 
between twisted and dual twisted cocycles, 
$\varphi_{L,R} \in H^n_{\pm \omega}$, defined as,
\bea
\langle \varphi_L \, , \varphi_R \rangle \equiv 
{1 \over (2 \pi i)^n} \int_\Sigma (u \, \varphi_L) \wedge (u^{-1} \, \varphi_R) \ ,
\label{eq:def:interxphiLphiR}
\eea
where $\Sigma$ is ${\mathbb C}^n$ deprived of the hypersurfaces corresponding to the poles of $\omega$.

The intersection number, therefore, plays the role of a {\it inner (scalar) product}, 
and the intersection matrix represents the {\it metric} of $H^n_{\pm \omega}$ 
\footnote{$\varphi_{L,R}$ in (\ref{eq:def:interxphiLphiR}) are understood as twisted cocycles and dual twisted cocycles with compact support. 
Moreover, let us recall that 
the decomposition problem can be addressed in the homology space, so that the integral decomposition stems from the decomposition of twisted cycles in terms of master cycles \cite{Mastrolia:2018uzb}.}.

\section{Applications} \label{sec:application}

The decomposition algorithm {\it via} intersection numbers discussed in the previous section 
can be applied to investigate the algebraic structure of any class of integral functions of the type (\ref{eq:def:twistedintegral}). 
In the following, 
we present its applications to integrals relevant to computational Quantum Mechanics and Quantum
Field Theory,
showing that the algebra of orthogonal polynomials, 
of the matrix elements, and of Green's functions,
can be derived within twisted de Rahm theory.
To make manifest the role of intersection theory within these contexts, we follow a simple strategy, 
based on mapping the integrals of the problems to univariate integral representations of type (\ref{eq:def:twistedintegral}),
and deriving the algebraic relations they obey from the master decomposition formula (\ref{eq:masterdecoformula}).

For each case, we systematically apply the following evaluation algorithm: 

\

\noindent
{\bf 1.} 
Identify a univariate twisted period integral of the form (\ref{eq:def:twistedintegral}),
\bea
\int_\Gamma \mu \, \varphi \ ,
\label{eq:def:mutwistedintegral}
\eea
characterized by the twist $\mu$, and by the univariate twisted cocycle $\varphi = {\hat \varphi} \, dz$. 

\

\noindent
{\bf 2.}
If $\mu$ is not multivalued, replace it with the regulated twist $u=u(\rho)$, 
by introducing a regulator $\rho$, so that, for a suitable value $\rho_0$ of the regulator, $u(\rho_0) = \mu$,
and determine the dimension of the cohomology group $\nu = {\rm dim}(H^1_\omega)$. 

\

\noindent
{\bf 3.}
Choose the cocycle basis 
      $e_i \equiv {\hat e}_i \, dz$  
      $\in H_\omega^1$,  
      with the first element having ${\hat e}_1 = 1$. \\
Choose the dual basis $h_i \equiv {\hat h}_i \, dz $ 
      $\in H_{-\omega}^1$ 
      with 
      ${\hat h}_i={\hat e}_i$,
      therefore implying ${\bf C}_{ij} = \langle e_i , e_j \rangle$. 
      
\

\noindent
{\bf 4.}
Decompose $\varphi$ in terms of the basis $e_i$, by using  
the master decomposition formula (\ref{eq:masterdecoformula}),
and determine the coefficients of the decomposition $c_i$ {\it via} intersection numbers, 
\bea
\varphi &=& c_1 \, e_1 + c_2 \, e_2 + \ldots + c_\nu \, e_\nu \ , 
\eea
(by taking, eventually, the $\rho \to \rho_0$ limit, to remove the dependence on the regulator). \\

\noindent
{\bf 5.}
Finally, translate the cocycle decomposition into integral decomposition, 
\bea
\int_\Gamma \mu \, \varphi &=& c_1 \, E_1 + c_2 \, E_2 + \ldots + c_\nu \, E_\nu \ , 
\eea
with
\bea
E_1 \equiv \int_\Gamma \mu \, dz \ , \quad {\rm and} \ E_i = \int_\Gamma \mu \, e_i \ , (i \ne 1) \ ,
\eea
and compare the result with the literature \footnote{The choice of the basis $e_i$ is arbitrary, 
and the result $\sum_i c_i E_i$ does not depend on the choice of $E_i$).
We opt for $e_1 = dz$ for illustration purposes.
Alternative choices, such as $e_1 = d\ln(z)$ are equivalent. 
In general, intersection numbers of twisted ${\rm d}\log$-forms can be directly computed by means of (global) Residue theorem \cite{Mizera:2019gea} - an aspect we will elaborate on, in the context of Quantum Mechanics, elsewhere.}.

Let us observe that: 
{\it i.} if $\nu=1$, the result comes just from the contribution of $E_1$; 
{\it ii.} if $\nu > 1$ and $c_i = 0$ for $i > 1$, the result still comes just from the contribution of $E_1$; 
{\it iii.} given its definition, $E_1$ represents the total volume of the integration domain. \

\subsection{Orthogonal Polynomials}

Univariate orthogonal polynomials $P_n=P_n(z)$ over an integration interval say $\Gamma$, 
labelled by integer indices $n$ are known to obey orthogonality conditions generically expressed as
\bea
\int_\Gamma 
\, \mu \, P_n P_m \, dz  = f_{n} \, \delta_{nm} \ , 
\label{eq:orthogonality}
\eea
which can be naturally cast in the form (\ref{eq:def:mutwistedintegral}), by simply interpreting
\bea
\label{eq:neworthogonalityaux1}
\varphi &\equiv& P_n \, P_m \, dz \ ,
\eea
as twisted cocycle.

We apply our evaluation algorithm to the set of orthogonal polynomials listed in Table \ref{tab:ops},
demonstrating that the orthogonality relation (\ref{eq:orthogonality}) emerges from the decomposition formula,
\bea
\int_\Gamma 
\, \mu \, \varphi  = c_1 \, E_1 \ .
\eea

For each polynomial type, in Tab. {\ref{tab:ops}}, we provide the relevant data needed for the decomposition via intersection numbers:
the regulated twist $u$, the functions ${\hat e_i}$ characterizing the bases of twisted cocycles;
the $\bf C$ matrix (eventually including the $\rho$ dependence), 
the value of $\rho_0$; the expression of $E_1$ and of $c_1$ 
yielding agreement with the results known in the literature.

Let us observe that, given the expression of $u$, in the case of Hermite polynomials, the vector space dimension is $\nu=2$, 
yielding $\varphi = c_1 \, e_1 + c_2 \, e_2$, with $c_2 = 0$. 

Moreover, in the case of the Laguerre and of the Gegenbauer polynomials, the integration measure $\mu$ and the twist $u$ coincide, therefore the coefficients $c_1$ and $E_1$ are exact in $\rho$, and no limit on $\rho$ is required.

\begin{widetext}
\begin{center}
\begin{table}[!h]
    \centering
    \begin{tabular}{|c||c|c|c|c|c|c|c|}
       \hline
       {\rm type} & $u$ & $\nu$ & ${\hat e}_i$ 
       & $\bf C$-matrix & $\rho_0$ & $E_1$ & $c_1$\\
       \hline 
       \hline
       \hline
         Laguerre $L_n^{(\rho)}$ & $z^\rho \exp({-z})$ & 1 & 1 
         & $\rho$ 
         & -- 
         & $\Gamma(1+\rho)$  
         & 
         $(\rho+1) (\rho+2) \cdots (\rho+n)/n!$  
         \\ 
       \hline
         Legendre $P_n$ & $(z^2-1)^\rho$ & 1 & 1 
         & $2\rho/(4 \rho^2-1)$ & 0 & 2 
         & $1/(2n+1)$ \\ 
       \hline
         Tchebyshev $T_n$ & $(1-z^2)^\rho$ & 1 & 1 
         & $2\rho/(4 \rho^2-1)$ & $-1/2$ & $\pi$ 
         & $1/2$ \\ 
       \hline
         Gegenbauer $C_n^{(\rho)}$ & $(1-z^2)^{\rho-1/2}$ & 1 & 1 
         & $(1-2\rho)/(4 \rho (\rho-1))$ 
         & -- & $\sqrt{\pi} \Gamma(1/2+\rho)/\Gamma(1+\rho)$ 
         & 
         $\rho ( 2\rho (2\rho+1) \cdots (2\rho+n-1) )/((n+\rho) n!)$ 
         \\ 
       \hline
         Hermite $H_n$ & $z^\rho \exp({-z^2})$ & 2 & $1, 1/z$ 
         & diagonal$(1/2,1/\rho)$ & 0 & $\sqrt{\pi}$ 
         & $2^n n!$  \\ 
        \hline
    \end{tabular}
    \caption{Functions and parameters of the decompositions involving orthogonal polynomials}
    \label{tab:ops}
\end{table}
\end{center}
\end{widetext}

\subsubsection*{Harmonic Oscillator and Hydrogen atom}

The computations we have just done can be easily extended to the computation of the matrix elements of powers of operators in position space, for instance. 
We illustrate some examples involving powers of 
the position operator, i.e.
$\langle \bullet | z^k | \bullet \rangle$, 
where $k$ may be a positive or negative integer, 
for two celebrated physics cases, the harmonic oscillator and the Hydrogen atom, in Quantum Mechanics. \\

\paragraph{Harmonic Oscillator.}

The eigenfunctions of the unidimensional Harmonic Oscillator 
in position space ($x \equiv z$), with principal quantum number $n$, 
(for unitary mass and pulsation, $m=1=\omega$)
are defined as

\bea
  \langle z | n \rangle &=& \psi_n(z) =
e^{-\frac {z^2}{2}} \, W_n(z) \ , \\
W_n(z) &\equiv& N_n \, H_n(z) 
\eea
in terms of Hermite polynomials,
where the normalization factors are,
\bea
    N_n \equiv (2^n n! \sqrt \pi)^{-\frac 12} \ .
\eea
The matrix elements $\langle m | z^k | n \rangle$ can be cast in the form (\ref{eq:def:mutwistedintegral}) as,
\bea
\langle m | z^k | n \rangle = 
\int_{-\infty}^{\infty} 
\!\! dz  \, \psi_m(z) \, z^k \, \psi_n(z) 
= \int_\Gamma \mu \, \varphi \ ,
\label{eq:def:HOintegrals}
\eea
with 
\bea
\mu &\equiv& e^{-{z^2}} \ , \\
\varphi &=& W_m(z)\, z^k \, W_n(z) \, dz \ .
\eea

According to our evaluation algorithm,
we decompose $\varphi$ using the master decomposition formula (\ref{eq:masterdecoformula}),
by following the same pattern previously applied to Hermite's polynomials (see Tab. \ref{tab:ops}),
yielding,
\bea
\int_\Gamma \mu \, \varphi 
&=& c_1 \, E_1 \ . 
\eea

In Tab. {\ref{tab:qm}}, we summarize the relevant ingredients of the decomposition.
They can be used to test our algorithm, and reproduce the following known cases:
\bea
 \langle n|m \rangle &=& \delta_{nm} \ , \\
 \langle n|z^{2k+1}|n\rangle &=&0 \ , \\
 \langle n|z^4|n\rangle &=&\frac 34 (2n^2+2n+1) \ , \\
 \langle n|z^3|n-3\rangle &=&\sqrt{n(n-1)(n-2)/8} \ ,\\ 
 \langle n|z^3|n-1 \rangle &=&\sqrt{9n^3/8} \ .
\eea

The mean values of the Hamiltonian operator $\langle n | H | n \rangle$, 
with $H$ in coordinate space, defined as $H \equiv (1/2) ( - \nabla^2 + z^2)$,
yield twisted period integrals with $\varphi$ being a linear combination of even powers of $z$, i.e.
$\varphi = \sum_{k=0}^n b_k \, z^{2k}$, for suitable coefficients $b_k$. 
We verified that their decomposition via intersection numbers give the expected result 
$\langle n | H | n \rangle = (n + 1/2).$ \\ 

\paragraph{Hydrogen Atom.}

The radial eigenfunctions of the H-atom 
in position space ($r=z$),
with principal quantum number $n$,
and orbital quantum number $\ell$,
(for unitary Bohr radius $a_0=1$)
are defined as
\bea
  \langle z | n,\ell \rangle &=& R_{n,\ell}(z) 
=
e^{-\frac {z}{n}} \,
W_{n,\ell}(z) \ , \\
W_{n,\ell}(z) &\equiv& 
  N_{n \ell} 
  \, \left({2z \over n}\right)^{\ell} \, 
  L^{2\ell+1}_{(n-\ell-1)}\bigg({2 z \over n}\bigg) \ , \quad 
\eea
in terms of Laguerre polynomials,
where the normalization factors are,
\bea
N_{n \ell}  &=& 
\left({2 \over n} \right)^{3/2} \, 
\sqrt{
(n-\ell-1)! 
\over 2 \, n \, (n+\ell)!
} \ ,
\eea

For illustration purposes, let us consider matrix elements for arbitrary principal quantum number $n$,
and identical orbital quantum number $\ell$, of the type,
\bea
\langle n_1, \ell | z^k | n_2, \ell \rangle = 
\int_{0}^{\infty} 
\!\!\! dz \, z^2 \, \, R_{n_1,\ell}(z) \, z^k \, R_{n_2,\ell}(z) \ , \quad 
\eea
which can be cast in the form (\ref{eq:def:mutwistedintegral}) as,
\bea
\langle n_1, \ell | z^k | n_2, \ell \rangle = \int_\Gamma \mu \, \varphi
\label{eq:def:HAtomintegrals}
\eea
with 
\bea
\mu &\equiv& z^2 \, e^{- z \left({1 \over n_1} + {1 \over n_2} \right)} \ , \\
     \varphi &=& W_{n_1, \ell}(z) \, z^k \, W_{n_2, \ell}(z) \ .
\eea

According to our evaluation algorithm,
we decompose $\varphi$ using the master decomposition formula (\ref{eq:masterdecoformula}),
by following a pattern which is similar to the one applied to the Laguerre polynomials, yielding,
\bea
\int_\Gamma \mu \, \varphi 
&=& c_1 \, E_1 \ . 
\eea

Tab. {\ref{tab:qm}} contains the relevant ingredients of the decomposition.
They can be used to test our algorithm, and reproduce the following known cases:

\bea
    \langle n_1,\ell| n_2, \ell \rangle&=&\delta_{n_1 n_2} \ ,\\
    \langle n,\ell| z | n, \ell \rangle&=&\frac 12 [3n^2-\ell(\ell+1)] \ ,\\
    \langle n,\ell| z^{-1} | n,\ell \rangle&=&\frac 1{n^2} \ , 
\eea

\bea
\langle n,\ell| z^{-2} | n,\ell \rangle&=&\frac 2{n^3(2\ell+1)} \ , \\
    \langle n,\ell| z^{-3} | n,\ell \rangle&=&\frac 2{n^3 \ell(\ell+1)(2\ell+1)} \ .
\eea

\noindent

\begin{widetext}
\begin{center}
\begin{table}[!ht]
    \centering
    \begin{tabular}{|c||c|c|c|c|c|c|}
       \hline
       {\rm type} & $u$ & $\nu$ & ${\hat e}_i$ 
       & $\bf C$-matrix & $\rho_0$ & $E_1$ \\
       \hline 
       \hline
       \hline
         Harmonic Oscillator $W_n$ 
         & $z^\rho \exp({-z^2})$ 
         & 2 
         & $1, 1/z$ 
         & diagonal$(1/2,1/\rho)$ 
         & 0 
         & $\sqrt{\pi}$ 
           \\ 
       \hline
       H-atom $W_{n,\ell}$ 
       & $z^\rho \exp({-z})$ 
       & 1 
       & 1 
       & $(n_1 n_2/(n_1+n_2))^2 (2+\rho)$ 
       & 0 
       & $2 (n_1 n_2/(n_1+n_2))^3$    \\ 
        \hline
    \end{tabular}
    \caption{Functions and parameters of the decomposition involving eigenfunctions.}
    \label{tab:qm}
\end{table}
\end{center}
\end{widetext}

\subsection{$n$-point Green's functions}

The Euclidean $n$-point Green's functions in Field Theory, $G_n=G_n(x_1,\ldots,x_n)$ is a 
for generic fields $\phi(x)$, 
and for any given action $S_E$, is defined as 
\bea
G_n &\equiv& { 
\int {\cal D}\phi \, \phi(x_1) \cdots \phi(x_n) \, e^{-S_E}
\over 
\int {\cal D}\phi \, \, e^{-S_E} \ 
} \ .
\eea
This equation is equivalent to,
\bea
\int {\cal D}\phi \, \phi(x_1) \cdots \phi(x_n) \, e^{-S_E} = G_n \int {\cal D}\phi \, \, e^{-S_E} \ ,
\eea
which can be read as a relation between integral of type (\ref{eq:def:mutwistedintegral}), 
\bea
\int_\Gamma \mu \, \varphi = G_n \, E_1 \ ,
\label{eq:genericdeco}
\eea
upon defining,
\bea
\mu &\equiv& e^{-S_E} \ , \\
\varphi &\equiv&   \phi(x_1) \cdots \phi(x_n) \, {\cal D}\phi \ , \\
E_1 &\equiv& \int_\Gamma \mu \, e_1 \ , \quad {\rm with} \ e_1 = {\cal D}\phi \ .
\eea
Therefore,
$G_n$ can be interpreted as the coefficient of the projection of the cocycle $\varphi$ on the master form $e_1$,
i.e. $\varphi = c_1 \, e_1$, with $c_1 = G_n$, and it can be determined within intersection theory, as observed in  \cite{Weinzierl:2020nhw,Gasparotto:2022mmp}.

\subsubsection{Single field, $\phi^4$-theory}

Let us consider a toy theory for a real scalar field $\phi(x)$, defined by the action
\bea
S_E &\equiv& S_0 + \epsilon S_1 \ , \\
 {\rm with} && S_0 = {1 \over 2} \gamma \, \phi^2(x) \ , \quad
S_1 = \, \phi^4(x) \ ,  
\eea
where $S_0$ represents the free kinetic term, and $S_1$, a quartic self-interaction term, with coupling constant $\epsilon$. 

By replacing $\phi(x)$ with the coordinate $z$, i.e. $\phi(x) \equiv z$,
the $n$-point Green's function $G_{n}$ for this theory can be defined through (\ref{eq:genericdeco}),
and can be determined by applying our computation algorithm to the decomposition of the cocycle $\varphi$, 
\bea
\varphi = z^{n} \, dz \ .
\eea

\paragraph{Free theory.}

The $n$-point Green's function $G_n^{(0)}$ in the free theory, 
is defined by considering just the kinetic term in the definition of 
\bea
\mu \equiv e^{-S_0} \ ,
\eea
and it can be computed 
by using the master decomposition formula (\ref{eq:masterdecoformula}).
In fact, as for the case of the Hermite polynomials, let us consider $u$ defined as
\bea
u \equiv z^\rho \, \mu  \ ,
\eea
such that $\lim_{\rho \to 0} u = \mu$. 
For this type of Gaussian integrals, the dimension of the cohomology group is $\nu = 2$, 
(see the case of Hermite polynomials in Tab.{\ref{tab:ops}}, which can be obtained by setting $\gamma=2$),
and we take the following basis of cocycles and dual cocycles, 
$\{{\hat e}_1,{\hat e}_2\}$ 
= $\{{\hat h}_1,{\hat h}_2\}$ 
= $\{1, 1/z\}$, 
yielding the intersection matrix,
\bea
{\bf C} = 
\left(
\begin{array}{cc}
     {1 \over \gamma }& 0   \\
    0 & {1 \over \rho }
\end{array}
\right).
\eea
By applying the master decomposition formula (\ref{eq:masterdecoformula}), 
and taking the $\rho \to 0$ limit, 
the decomposition of $\varphi$ in terms of the master forms $e_1$ and $e_2$, reads
$\varphi = c_1 \, e_1 + c_2 \, e_2$, with $c_2 = 0$, and 
\bea
c_1 = G_{n}^{(0)} = {1 \over \gamma^{n/2}} (n-1)!! \ , \quad {\rm for \ even \ } n \ .
\label{eq:npointG0funtion}
\eea
This result corresponds to the application of Wick's theorem in Quantum Field Theory, which in the free theory allows to rewrite any $n$ point functions
combinatorially in terms of products of two point functions.
From the general result of the $n$-point correlator, 
we can read the 2-point correlation function for the free theory, corresponding to the propagator of the $\phi$ field, 
\bea
G_{2}^{(0)} = {1 \over \gamma} \ .
\eea

\paragraph{Perturbation Theory.}
The $n$-point correlation function $G_n$ in the full theory can be computed perturbatively, in the small coupling limit, 
$\epsilon \to 0$, and expressed in terms of $G_n^{(0)}$. 

Let us describe the determination of the next-to-leading order (NLO) corrections to the 2-point function,
\bea
G_2 &=& { \int dz \ z^2 \ e^{-S_0 - \epsilon S_1} 
          \over 
        \int dz \ e^{-S_0 - \epsilon S_1} } \nonumber \\
&=& 
{ 
\int dz \ z^2 \ e^{-S_0} (1 - \epsilon S_1 + \ldots) 
\over 
\int dz \ e^{-S_0} (1 - \epsilon S_1 + \ldots)
} \nonumber \\
&=& 
{
\Big(
G_2^{(0)} - \epsilon \, G_6^{(0)} + \ldots 
\Big) 
\Big(
1 + \epsilon \, G_4^{(0)} + \ldots 
\Big)
}
\nonumber \\
&=& 
G_2^{(0)} + \epsilon \Big(
G_2^{(0)} G_4^{(0)} - G_6^{(0)}
\Big) + {\cal O}(\epsilon^2) 
\nonumber \\
&=& {1 \over \gamma} \bigg(1 - 12 \epsilon {1 \over \gamma^2} \bigg) + {\cal O}(\epsilon^2) \ ,
\eea
where the term proportional to $\epsilon$ is the NLO correction to the free propagator. Notice that in this result Wick's theorem still appears in the combinatorics 
of the $G_{2j}^{(0)}$ terms.

\paragraph{Exact theory.}

Let us consider now the decomposition of $\varphi = z^{n} \, dz$ in the exact theory, with 
\bea
\mu \equiv e^{-S_E} \ ,  
\eea
and evaluate the intersection numbers with
\bea
u \equiv z^\rho \, \mu \ ,
\eea
such that $\lim_{\rho \to 0} u = \mu$. 
In this case, 
$\nu = 4$, 
namely, the dimension of the twisted cohomology group is larger than in the free theory case.
We choose a basis of cocycles, 
$\{{\hat e}_1,{\hat e}_2,{\hat e}_3,{\hat e}_4\}$ 
= $\{1, 1/z, z, z^2\}$, 
and for the dual cocycles $ \{{\hat h}_i\}_{i=1}^4 = \{{\hat e}_i\}_{i=1}^4 $,
yielding the intersection matrix,
\bea
{\bf C} = 
\left(
\begin{array}{cccc}
     0 & 0 & 0 & {1 \over 4 \gamma }   \\
     0 & {1 \over \rho }& 0 & 0 \\
     0 & 0 & {1 \over 4 \gamma } & 0 \\
     {1 \over 4 \gamma } & 0 & 0 & - {\gamma \over 16 \epsilon^2 }
\end{array}
\right).
\eea

The master decomposition formula (\ref{eq:masterdecoformula}) can be used to project the cocycle 
$\varphi = z^4 \, dz$ onto the master forms, which, after taking the $\rho \to 0$ limit reads,
\bea
\varphi = 
c_1 \, e_1 + c_2 \, e_2 + c_3 \, e_3 + c_4 \, e_4 \ ,
\eea
with
\bea
c_1 = {1 \over 4 \epsilon} \ , \quad 
c_2 = 0 \ , \quad 
c_3 = 0 \ ,
\qquad c_4 = - {\gamma \over 4 \epsilon} \ .
\eea
The cocycles decomposition translates into the integral relation,
\bea
\int_\Gamma dz \, z^4 \, e^{-S_E} = 
c_1 \int_\Gamma dz \, e^{-S_E} + c_4 \int_\Gamma dz \, z^{2} \, e^{-S_E} \ ,
\eea
which, by dividing both sides by the first integral appearing in the {\it r.h.s.}, 
can be rewritten as a relation between $n$-point functions in the exact theory, 
\bea
%
G_4 &=& c_1 + c_4 G_2 \ .
\eea

This relation can be used to express $G_2$ in terms of $G_4$, as 
\bea
G_2 &=& 
{1 \over \gamma} \Big( 1 - 4 \epsilon G_4 \Big) \ ,
\eea
which is an all-order result.
To verify that it is compatible with the result earlier obtained in perturbation theory, 
we observe that in order to determine $G_2$ up to the first order in $\epsilon$, 
it is sufficient to keep just the leading order of $G_4$ in the {\it r.h.s.}
i.e.
$G_4 = G_4^{(0)} + {\cal O}(\epsilon)$,
implying 
\bea
G_2 &=&
{1 \over \gamma} \Big( 1 - 4 \epsilon G_4^{(0)} \Big) 
+ {\cal O}(\epsilon^4),
\eea
which corresponds to the same result obtained in perturbation theory,
upon substituting 
\bea
G_4^{(0)} = {1 \over \gamma^2} 3!! = {3 \over \gamma^2} \ ,
\eea
which can be read out of (\ref{eq:npointG0funtion}).

\section*{Conclusion} \label{sec:conclusion}
We showed that Intersection Theory of Twisted de Rham Cohomolgies plays a pivotal 
role on the algebraic structure of special functions that appear 
in Quantum Mechanics and Quantum Field Theory.

We applied de Rham's theory to simple, univariate integrals, built out of orthogonal polynomials, 
quantum mechanical eigenfunctions, and fields, by interpreting them as twisted period integrals, 
namely as pairings of twisted cycles and cocycles.
We derived the algebraic properties of these integral functions from the decomposition properties of cocycles, 
showing that the linear relations they obey can be derived by means of intersection numbers of twisted cocycles.
We exported to Quantum Mechanics and Quantum Field Theory 
a decomposition algorithm and computational tools recently developed in the context of scattering amplitudes' and Feynman integrals' calculus.
Orthogonality relations for polynomials, 
matrix elements of quantum mechanical operators, 
Green functions within Wick's theorem (hence moments of distributions within Isserlis' theorem), 
therefore provided additional proof of evidence on the role of the de Rham Intersection Theory in fundamental physics.

When these analyses developed within physics contexts are combined with the study of Aomoto-Gel'fand integrals, Euler-Mellin integrals, 
GKZ hypergeometric systems, and other special functions, which have been the natural target of investigation within pure mathematical areas, like differential and algebraic topology, combinatorics and number theory,
they seem to point toward a uniform framework ruling calculus which spans across various scientific disciplines.

Our results are applicable to the study of generalised moments of probability distributions:
the dimension of the cohomology groups corresponds to the number of independent moments - which we can called master moments;
the intersection numbers for twisted cocycles can be used to derive linear and quadratic relations among them. 
The latter can be used for decomposing all moments of a distribution in terms of master moments, as well as to build functional equations, difference and differential equations to evaluate the master moments.
Regulated twists allow considering moments of distributions with generic powers.
In this fashion, one can export the experience and tools developed in Feynman calculus in all problems that admit a statistical interpretation.

In our vision, the moments of distribution admit a physical interpretation in terms of generalised fluxes and, therefore represent conserved quantities, invariant under deformations of the twisted cocycles: 
De Rham twisted theory = Stokes's theorem for fluxes of singular differential forms through hypersurfaces with holes. 

Our study shows that eigenfunctions in Quantum Mechanics and fields in Quantum Field Theory, which represent elements of Hilbert and Fock's spaces, respectively, that can have infinite dimensions, generate matrix-elements that belong to de Rham's cohomology groups, 
(or better to spaces that are isomorphic to them)
which have a finite number of dimensions. 
Therefore, the knowledge of the master moments and the inner product within the cohomology group is sufficient to determine all moments of distributions.

Our analysis, carried out for univariate integral representations, points to the importance of developing efficient methods for the computations of multivariate intersection numbers for twisted cocycles, which are crucial for dealing with moments of generical multivariate distributions.

The simple, paradigmatic applications discussed here can be considered the first step toward a systematic analysis of Quantum Mechanics in light of Intersection Theory. 
On the Quantum Field Theory [Statistics] side, 
exploiting the freedom of different choices of master moments, 
within the perturbative diagrammatic expansion ruled by Wick theorem [Isserlis's theorem], 
or directly in the exact theory of path integrals, 
may lead to equivalent representations of the S-matrix [partition function], 
yet expressed in terms of different sets of basic Green's functions [correlator functions], 
hopefully not diverging, 
which may offer new layouts for investigating the finiteness of physical theories [statistical models] around critical values of space-time dimensions [of parameters of the models]. 

We would like to conclude this study with the speculation that 
the invariance under twisted de Rham co-homology groups 
seems to extend and generalize the role of gauge invariance of matrix elements in Quantum Field Theory. 

We hope that our investigation could offer a new point of view on the intertwinement between physics, geometry, and statistics, 
which exploits the analogy between fluxes, period integrals, and statistical moments,
and which could suggest a general approach to be applied in several scientific contexts. \\ \\

\paragraph*{Notes.} During the completion of this work, an important study appeared \cite{Gasparotto:2022mmp}, 
which provides additional evidence on the role of cohomological methods and their interplay with computational techniques for Feynman integrals, in addressing relevant problems in perturbation theory as well as beyond it. \\ 

\paragraph*{Acknowledgements.} 
We thank Thibault Damour, Hjalte Frellesvig, Federico Gasparotto, and Nobuki Takayama for comments on the manuscript.

\bibliography{biblio.bib}


\end{document}